\documentclass[a4paper,10pt,twoside]{cpc-hepnp}

\usepackage{multicol}
\usepackage{graphicx}
\usepackage{booktabs}
\usepackage{amssymb,bm,mathrsfs,bbm,amscd}
\usepackage[tbtags]{amsmath}
\usepackage{lastpage}
\usepackage{CJK}

\begin{document}
\begin{CJK*}{UTF8}{gbsn}


\title{Finite-size behavior near the critical point of QCD
phase-transition\thanks{Supported in part by the National Natural Science Foundation of China with project No. 10835005, 11221504, and 11005046, and the MOE of China
for doctoral site with project No. 20120144110001. }}

\author{%
      Lizhu Chen$^{1;1)}$\email{chenlziopp@gmail.com}%
\quad Yunyun Chen$^{1}$
\quad Yuanfang Wu$^{2}$%
}
\maketitle

\setlength\unitlength{1cm}
\begin{picture}(0,0)
\put(12,3.2){\small Submitted to‘Chinese Physics C'}
\end{picture}

\address{%
$^1$ School of Physics and Optoelectronic Engineering,\\ Nanjing University of Information Science and Technology, Nanjing 210044, China\\
$^2$  Key Laboratory of Quark and Lepton Physics (MOE) and
Institute of Particle Physics, \\Central China Normal University, Wuhan 430079, China\\
}

\begin{abstract}
It is pointed out that finite-size effect is not negligible in
locating critical point of QCD phase transition at current
relativistic heavy ion collisions. The finite-size scaling form of
critical related observable is suggested. Its fixed point behavior
at critical incident energy can be served as a reliable
identification of critical point and nearby boundary of QCD phase
transition. How to find experimentally the fixed point behavior is
demonstrated by using 3D-Ising model as an example. The validity of
the method at finite detector acceptances at RHIC is also discussed.
\end{abstract}





\begin{multicols}{2}

\section{Introduction}

 Quantum colordynamics (QCD) has predicted quark
deconfinement and chiral symmetry restoration at finite temperature
and density~\cite{qgp}. Lattice-QCD has shown that the transition is
crossover at vanishing baryon chemical potential $\mu_{\rm
B}$~\cite{c-crossover}. The QCD based model indicates that the
crossover turns to be a first-order phase transition at larger
values of $\mu_{\rm B}$~\cite{1st}. The endpoint of the first order
phase transition to the crossover is referred as critical endpoint,
or critical point. All these show that the transition from hadron
phase to quark-gluon-plasma (QGP) phase can happen in 3 possible
ways.

The data from current relativistic heavy ion experiments show that
the QGP has been formed at RHIC energies~\cite{qgp-rhic}. But the
position of critical point in QCD phase diagram is not clear from
the theoretical side. Well defined character of critical point, the
divergence of correlation length, warrants the possibility of
finding it experimentally. The goal of the beam energy scan at RHIC,
and future heavy ion experiment at FAIR is aimed to pass through the
critical incident energy.

In relativistic heavy collision, both the size and duration of
formed system are finite. From event to event, the overlapping area,
i.e., the formed system size, varies with impact parameter. For
finite-size system, critical behavior changes with system size
($L$). If the system size is too small, the correlation length can
not be fully developed to cause a phase transition. If the system
size is large enough and the correlation length ($\xi$) is much
smaller in comparison with system size, the system can still be
considered as infinite large. The critical behavior under thermal
limit is available. This is why the non-monotonic behavior is
suggested as an indicator of critical
point~\cite{non-monotonic,cpsignal-1,Asakawa-3rd} in a long period.

Nevertheless, non-monotonic behavior is not unique to critical
point. In case of first order phase transition, or crossover, some
observables also show the non-monotonic behavior~\cite{c-crossover}.
The absence of non-monotonic behavior does not exclude the existence
of critical point, such as the maximum cluster size in 3D-Ising
model shown in Fig.~1(a).

Moreover, if the correlation length is comparable with system size,
the finite-size effect is not negligible. When correlation length is
large than $\frac{1}{6}$ of system size, it has been shown that the
finite-size effect has to be taken into
account~\cite{prl-fs,prb-fs}.

Although, it is still difficult to estimate the size of the formed
system and correlation length at critical point in relativistic
heavy ion collisions. A rough estimation shows that the system size
at freeze-out is less than 12fm~\cite{HBT-rhic,stephanov-PRD}. The
correlation length is round 6fm for typical nuclear
collisions~\cite{Antoniou,stephanov-PRD}. After considering the
finite evolution time, or finite-size, it is argued that the maximum
of correlation length may not be beyond 2-3fm at critical
point~\cite{rajagopal-ft,guy-fs,rajagopal-cpod-bnl}. Base on those
estimations, the ratio of correlation length to system size is round
$\frac{1}{6}$-$\frac{1}{2}$, which is in the region larger than
$\frac{1}{6}$. So in current relativistic heavy ion collisions at
RHIC, the finite-size effect most probably has to be taken into
account, rather than negligible~\cite{stephanov-PRD}.

In accounting for the finite-size effects, the critical behavior of
all suggested observables, such as the fluctuations and correlations
of transverse momentum, multiplicity, conserved
charges~\cite{corr-fluc}, and in particular, the higher order
moments of conserved quantities~\cite{h-moments}, should be re-examined
under the frame of finite-size behavior of critical point and nearby
boundary.

In this, we firstly discuss the finite-size behavior of the
critical point, the first order phase transition, and the crossover
in general. Secondly, we suggest the finite-size scaling form of the
critical related observable in relativistic heavy ion collisions.
Its fixed-point behavior at critical incident energy can be served
as a reliable identification of critical point and nearby boundary
of QCD phase transition. We demonstrate how to locate the
fixed point from experimental observable by using 3D-Ising model as
an example. Finally, the validity of the method at finite detector
acceptance at RHIC is discussed.

\section{The Behavior of Finite Size Scaling and Fixed Point}

For the second order phase transition, the critical behavior is well
described by finite-size scaling. It was firstly proposed from
phenomenological~\cite{fss-1} and
renormalization-group~\cite{fss-RG} theories, and was approved by
the Monte Carlo results of finite systems in different universal
classes~\cite{fss-2}. This scaling form not only describes the
behavior of the observables at different system sizes, but also
indicates the position of the critical point and the critical
exponents in infinite system. Therefore, from the finite-size
scaling of critical related observables, the position and critical
exponents of critical point can be precisely extracted. This has
been implemented in locating critical point of multi-fragmentation
nuclear liquid-gas phase transition~\cite{nucl-lg}.

In contrast to the critical point, the finite-size behavior of first
order phase transition has not been well understood in
general~\cite{binder-review}. But the finite-size scaling behavior
of first order phase transition is shown to correspond to so-called
discontinuity fixed points of the renormalization group
transformations, which are characterized by eigenvalue exponents
equal to the spacial dimension~\cite{RG-first}.  Consequently, the finite-size
scaling form pertains, and the scaling exponents are the spatial
dimension, in contrary to the critical exponents of critical point.
The phenomenological theory of finite-size scaling at first-order
phase transition is proposed by K. Binder and D.P. Landau, and it is found
to be in good agreement with Monte Carlo simulation
results~\cite{Binder-first}.

Different from the critical point and the first order phase
transition, at the crossover region, there is no singularity in all
kinds of observables. The observables are system size
independent~\cite{c-crossover,f-crossover}. But it should be noticed
that this holds only when the system size is not too small. When the
system size is very small and the finite correlation length is
comparable with the system size, the observables will become larger
and larger when the system size goes to smaller and smaller.

In heavy ion collisions, the critical related observables are
generally considered to be the fluctuations of conserved charges,
like baryon number, electric charge, and
strangeness~\cite{corr-fluc,h-moments,koch}. The incident energy and
centrality dependence of some related observables are fully
investigated in current heavy ion experiments~\cite{exp}.

Incident energy $\sqrt s $ is a controlling parameter, like
temperature $T$, or external field $h$ in thermodynamic systems. The
centrality, i.e., impact parameter, presents the overlapped area of
two incident nuclei. It directly related to the size of the formed
system, and randomly fluctuates from event to event.

Therefore,  the finite-size scaling in nuclear collisions can be generalized
as following. When the size of the formed matter $L$ is much larger
than the microscopic length scale (which is less than 1fm) and
incident energy is near the critical one $\sqrt s_{\rm c}$, the
critical related observable, e.g., $Q(\sqrt s,L)$ in general, can be
written in a finite-size scaling form~\cite{fss-1,fss-RG,fss-2},
\begin{equation}\label{efss}
 Q(\sqrt s,L)=L^{\lambda/\nu}F_Q(\tau L^{1/\nu}).
 \end{equation}
\noindent Where $\tau=(\sqrt s-\sqrt s_{\rm c})/\sqrt s_{\rm c}$ is
reduced incident energy. $\nu$ and $\lambda$ are the critical
exponents of the correlation length $\xi =\xi_0 \tau^{-\nu}$ and the
observable, respectively. They characterize the universal class of
the phase transition. Finite-size scaling indicates that the
observable at different system sizes can be re-scaled to an
identical scaling function $F_Q$ with scaled variable $\tau
L^{1/\nu}$.

At critical energy, $\sqrt s= \sqrt s_{\rm c}$, the scaled variable
($\tau L^{1/\nu}=0$) is independent of system size $L$, and the
scaling function becomes a constant,
\begin{equation}\label{efss-1}
F_Q (0) = Q(\sqrt s_{\rm c},L)L^{-\lambda/\nu}.
\end{equation}
\noindent It shows that the fluctuation of critical related
observable is self-similar at different size scales. In the case,
the energy dependence of the observable at various of system sizes
will intersect to this point, i.e., {\it fixed point}. The energy of
the fixed point indicates the critical incident energy. As an
example, we show in Fig.~1(b) the fixed point behavior of the
maximum cluster in 3D-Ising model, which is supposed to be the same
university of the de-confinement~\cite{3d-ising}. We can see that
the maximum cluster at different lattice sizes intersect exactly at
the fixed point, i.e., critical point.

Reversely, if we can find the fixed point from incident energy
dependence of properly scaled observable in heavy ion collisions,
which are measured at different centralities, i.e., system sizes, it
will indicate the existence of critical point.

\end{multicols}

\begin{figure*}
\begin{centering}
\includegraphics[width=5.4in]{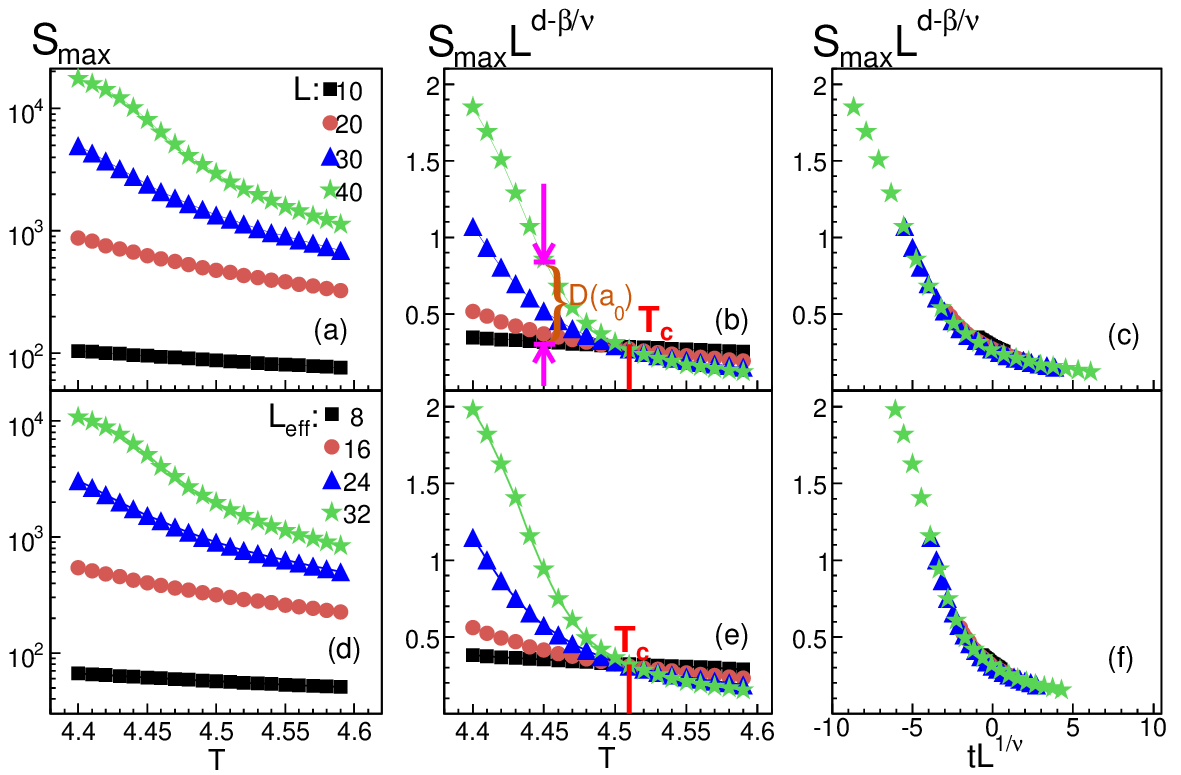}
\caption{\label{Fig. 2}(Color online) Upper panel: (a) and (b) are
the temperature dependence of maximum cluster size and maximum
cluster size scaled by $L^{d-\beta/\nu}$, and (c) is the scaling
function of maximum cluster size in 3-D Ising model at different
lattice sizes $L$. Lower panel: the same measures as in
corresponding upper panel, but in a sub-system containing 50\%  of the lattice sites.}
\end{centering}
\end{figure*}

\begin{multicols}{2}

In order to find the exponent of the scale, and incident energy of
the fixed point, we can firstly present the incident energy dependence
of critical related observable at different system sizes, similar to
Fig.~1(a). Then multiply a size factor to the observable $Q(\sqrt
s,L)$, i.e., $Q(\sqrt s,L)L^{-a}$, and change the parameter $a$ from
$-\infty$ to $\infty$ to see if all size curves interact to a point
for a certain value of $a_0$ at a certain incident energy, e.g.,
Fig.~1(b).

In experiment, the point liked behavior can be quantified by the
width of all size points. At a given incident energy, the width is
usually defined as the square root of $\chi^2$ of all size points,
i.e., 
\begin{eqnarray}\label{width-1}
 D(\sqrt s,a)=\sqrt{\frac{\chi^2_{Q(\sqrt s, L)L^{-a}}}{N_L-1}}.
\end{eqnarray}
\noindent
$N_L$ is the number of points, and $\chi^2_{Q(\sqrt s, L)L^{-a}}$ is
error weighted variance of all size points,

 \begin{eqnarray}\label{width-1}
\chi^2_{Q(\sqrt s, L)L^{-a}}=\sum_{i=1}^{N_L}\frac{\left[Q(\sqrt
s,L_i)L_i^{-a}-\langle Q(\sqrt s,L)L^{-a}\rangle\right]^2}{w_i^2}.
\end{eqnarray}\noindent

$w_i=\delta \left[Q(\sqrt s,L_i)L_i^{-a}\right]$
are the experimental error of $\left[Q(\sqrt s,L_i)L_i^{-a}\right]$,
where both the errors of the observable $Q(\sqrt s,L_i)$ and system
size $L_i^{-a}$ contribute to. $\langle Q(\sqrt s,L)L^{-a}\rangle$
is also error weighted mean,

\begin{eqnarray}\label{width-1} 
\langle Q(\sqrt s,L)L^{-a}\rangle=\frac{\sum_{i=1}^{N_L}Q(\sqrt s,L_i)L_i^{-a}/w_i^2}{\sum_{i=1}^{N_L} 1/w_i^2}.
\end{eqnarray} \noindent

For example, in Fig.~1(b), this width at a given temperature is the
distance between two violet arrows. Therefore, if at a given
incident energy, the minimum of $D(\sqrt s,a)$ is round 1 at $a_0$,
i.e., $ D_{\rm min}(\sqrt s,a_0) \sim 1$, it can be recognized as an
experimental point. While, if it keeps larger than 1, there is no
point liked behavior.

For QGP formed system~\cite{qgp-rhic}, the following 3 cases should
be expected. (1) $D_{\rm min}(\sqrt s,a_0)$ at a certain incident
energy is round 1, and at nearby incident energies, it is always
larger than 1, and corresponding $a_0$ is {\it not} an integer, as
green curve shown in the middle of Fig.~2. This may indicate the
existence of the fixed point, i.e., the critical point in
Eq.~(\ref{efss-1}). The critical incident energy is the energy of
the fixed point and the obtained parameter $a_0$ is the ratio of
critical exponents, i.e., $\lambda/\nu=a_0$, cf. Fig.~1(b).



In the case, the critical behavior should be further confirmed by
the scaling function,
\begin{equation}\label{efss-2}
F_Q(\tau L^{1/\nu})=L^{-a_0}Q(\sqrt s,L).
\end{equation}
\noindent Here the critical exponent of correlation length $\nu$ is
a fitting parameter. If the data at all incident energies and system
sizes can be well fitted by the scaling function, the critical point
and the critical exponents are finally determined, cf., fig.~1(c).

(2) $D_{\rm min}(\sqrt s, a_0)$ at a certain incident energy is
round 1, and at nearby incident energies, it is always larger than
1, and corresponding $a_0$ is an {\it integer}. This indicates also the existence of the
fixed point, but scaled power is trivial integer. It implies the
region of the first order phase transition. The incident energy of
the fixed point is the transition energy of the first order phase
transition. The scaling function of the observable should be simply
formulated by the spatial dimension, instead of the critical
exponents in Eq.~(1).

If $a_0$ is zero, there are two possibilities. It could be the
critical point with critical exponent $\lambda =0$, like Binder
cumulant ratio~\cite{binder1981}, or the region of the first order
phase transition. The final identification is their specified
scaling functions, as discussed above.

(3) $D_{\rm min}(\sqrt s, a_0)$ is round 1 at {\it all} incident
energies and
corresponding $a_0$ is an integer. This indicates all size curves
are overlapped, and there is in fact no fixed point. It corresponds
to the transition of crossover.

\section{Detector Effects}
It should be stressed that the observables mentioned here are the
intensive variable, like susceptibility. If the observables are
extensive variable, such as the fluctuation of the particle number,
$\langle(N-\overline{N})^2\rangle=TV \chi$, the trivial size
dependence are included, and can be merged to the power $a$.

The size of the formed matter in heavy collisions is mainly
determined by overlapping area of two incident nuclei. This area is
proportional to the number of participant nucleons and is quantified
as centrality. The initial size of the formed matter can be
approximately estimated by the square root of the number of
participants, $\sqrt {N_{\rm part}}$. The maximum size is $\sqrt
{2N_{\rm A}}$, $N_{\rm A}$ is the number of nucleons of incident
nucleus. The ratio,
 \begin{eqnarray}\label{el}
 L=\sqrt {N_{\rm part}}/\sqrt {2N_{\rm A}},
\end{eqnarray} \noindent 
presents the relative size of the initial system.

The system size $L^{\prime}$ at transition should be larger than
initial one $L$ and monotonically increase function of $L$, i.e.,
$L^{\prime}=cL^{1+\delta}$ with $\delta \ge 0$ in general. Whether
we take $L^{\prime}$ or $L$ in Eq.~(\ref{efss}), the scaling
exponents will be different, but the position of critical point will
be the same. So the initial size is a good approximation in locating
the position of critical point.

It should also be noticed that the detectors at current relativistic
heavy ion experiments cover a part of the phase space, and only a
part of final state particles is accepted. Even if the critical
related information are survived in the final state observables,
whether the finite-size behavior of detected subsystem is preserved
has to be examined further.

The finite size behavior of a sub-systems is demonstrated in 3-D
Ising model. The size of sub-system is chosen to be a certain
percent of the whole lattice sites. Changing the lattice of the
whole system, the effective sites of the sub-system, $L_{eff}$, vary
with it. We find that the finite size behavior of sub-system keeps
valid as long as the size of sub-system is within the range of
finite size scaling.

In the lower panel of Fig.~1, the finite size behavior of the
maximum cluster size at various $L_{eff}$ are presented. Where the
size of sub-system is $50\%$ of the whole system. In comparison with
the corresponding results of the whole system shown in upper panel
of Fig.~1, the susceptibilities of the sub-system is different from
that of the whole one, but the position of fixed point indicates the
same critical temperature, $T_{\rm c}=4.51J$. Moreover, the maximum
cluster at different sub-system sizes are well scaled to an
identical scaling function. Therefore, the suggested finite size
behavior should be visible at a detector with a relative large
acceptance, like RHIC/STAR.

\section{Summary}
In the summary, It is pointed out that the finite-size effects are not
negligible in locating the critical point of QCD phase transition at
current relativistic heavy ion collisions. At the crossover,
critical point and first order QCD phase transition, the finite-size
scaling behaviors of the critical related observable are suggested.

The critical point of QCD phase transition can be found by the
appearance of the fixed point with a non-integer power in scaled
size factor, and the finite-size scaling function of the observable.
The region of the first order phase transition is identified by the
fixed point with an integer power in scaled size factor and the
scaling function which is determined by spatial dimension.

At the region of the crossover, the behavior of the fixed point is
absent, and the scaling function reduces to the incident energy
dependence of observable, which is system size independent.

At a given incident energy, the width of the observables at various
centralities is suggested as a quantification of point liked
behavior. The energy dependence of the width at different orders of
phase transitions are shown. When incident energy scans from high to
low, the deviation of minimum width from point like behavior will
indicate the appearance of the critical point.

Finally, for a finite acceptance detector, we demonstrate that the
finite-size behavior of critical related observables keep valid as
long as the detected subsystem is large enough.

\end{multicols}

\vspace{-1mm}
\centerline{\rule{80mm}{0.1pt}}
\vspace{2mm}

\begin{multicols}{2}

\end{multicols}

\clearpage
\end{CJK*}

\begin{thebibliography}{90}

\vspace{3mm}

\bibitem{qgp}J. C. Collins, M. J. Perry, Phys. Rev. Lett, 1975,  {\bf 34}: 1353;
B. A. Freedman, L. D. Lerran, Phys. Rev. D, 1977, {\bf 16}: 1196; E. V.
Shuryak, Phys. Lett. B, 1981,  {\bf 107}: 103.

\bibitem{c-crossover}Y. Aoki, G. Endrodi, Z. Fodor, S. D. Katz, K.K. Szabo, Nature, 2006, 
{\bf 443}: 675; Y. Aoki, Z. Fodor, S.D. Katz, K.K. Szabo, Phys.
Lett. B, 2006, {\bf 643}: 46.

\bibitem{1st}Z. Fodor and S. D. Katz, JHEP, 2004,  {\bf 04}: 050;
Z. Fodor, S.D. Katz, and K. K. Szabo, Phys. Lett. B, 2003, {\bf 568}:73.

\bibitem{qgp-rhic}M. Gyulassy, L. McLerran, Nucl. Phys. A, 2005,  {\bf 750}:30-63; 
J.Adams, et al, (STAR Coll.), Nucl.Phys. A, 2005, {\bf 757}: 102.

\bibitem{non-monotonic}M. A. Stephanov, hep-ph/0402115, Int. J. Mod. Phys. A, 2005, 
{\bf 20}: 4387.

\bibitem{cpsignal-1}M. A. Stephanov, Phys. Rev. Lett. 2009,  {\bf 102}: 032301.

\bibitem{Asakawa-3rd}M. Asakawa, S. Ejiri, M. Kitazawa, Phys. Rev. Lett. 2009,  {\bf 103}: 262301.

\bibitem{prl-fs}C. Weber, L. Capriotti, G. Misguich, F. Becca,
M. Elhajal, and F. Mila, Phys. Rev. Lett. 2003, {\bf 91}: 177202.

\bibitem{prb-fs}P. Olsson, Phys. Rev. B, 1997,  {\bf 55}:3583.

\bibitem{HBT-rhic}J. Adams, et. al. (STAR Collaboration), Phys. Rev. Lett., 2004, {\bf 93}: 012301; S. S. Adler, et,
al. (PHENIX collaboration), Phys. Rev. Lett., 2004, {\bf 93}: 152302; R. A.
Soltz, J. Phys. G: Nucl. Part. Phys., 2005,  {\bf 31}: S325.

\bibitem{stephanov-PRD}M. A. Stephanov, K. Rajagopal, E. Shuryak, Phys. Rev. D {\bf 60} (1999) 114028.

\bibitem{Antoniou}N. G. Antoniou, F. K. Diakonos, and A. S.
Kapoyannis, Phys. Rev. Lett. {\bf 97} (2006) 032002.

\bibitem{rajagopal-ft}B. Berdnikov and K. Rajagopal, Phys. Rev. D, 2000,  {\bf 61}: 105017.

\bibitem{guy-fs}K. Paech, Eur. Phys. J. C, 2004,  {\bf 33}: S627.

\bibitem{rajagopal-cpod-bnl}K. Rajagopal, summary talk at 5th International Workshop on
Critical Point and Onset of Deconfinement at BNL, in June 8-12,
2009.

\bibitem{corr-fluc}M. A. Stephanov, K.
Rajagopal, and E. Shuyak, Phys. Rev. Lett., 1998,  {\bf 81}: 4816;  H.
Heiselberg, Phys. Rept., 2001,  {\bf 351}: 161.

\bibitem{h-moments}M. A. Stephanov, Phys. Rev. Lett., 2009, {\bf 102}: 032301;
Masayuki Asakawa, Shinji Ejiri, Masakiyo Kitazawa,
nucl-th 0904.2089.

\bibitem{fss-1}M. E. Fisher, in Critical Phenomena, Proceedings of the
International School of Physics Enrico Fermi, Course 51,
edited by M. S. Green (Academic, New York, 1971).

\bibitem{fss-RG}E. Br\'ezin, J. Phys., 1982,  {\bf 43}: 15.

\bibitem{fss-2}X. S. Chen, V. Dohm, and A. L. Talapov, Physica A, 1996, 
{\bf 232}: 375; X. S. Chen, V. Dohm, and N. Schultka, Phys. Rev.
Lett., 1996,  {\bf 77}: 641.

\bibitem{nucl-lg}M. K. Berkenbusch, {\it et al.}, Phys. Rev. Lett., 2001, {\bf 88}: 022701.

\bibitem{binder-review} K. Binder, Rep. Prog. Phys., 1987,  {\bf 50}: 783.

\bibitem{RG-first}J. M. J. van Leeuwen, Phys. Rev. Lett., 1975, 
{\bf 34}: 1056; B. Nienhuis and M. Nauenberg, ibid., 1975,  {\bf 35}: 477;
M.E. Fisher and A. N. Berker, Phys. Rev. B, 1982,  {\bf 26}: 2507.

\bibitem{Binder-first}K. Binder and D. P. Landau, Phys. Rev. B, 1984, {\bf 30}: 1477.

\bibitem{f-crossover}A. Ukawa, Lecture on Lattice QCD at finite
temperature (1993).

\bibitem{koch}S. Jeon and V. Koch, Phys. Rev. Lett., 2000, {\bf 85}: 2076;
V. Koch, arXiv: 0810.2520.

\bibitem{exp}J. Adams, et. al.(STAR
collaboration), Phys. Rev. C, 2005,  {\bf 72}: 044902; B. I. Abelev, et. al.
(STAR collaboration), Phys. Rev. C, 2009, {\bf 79}, 024906.

\bibitem{3d-ising}J. Garc\'a, J. A. Gonzalo, Physica A, 2003,  {\bf 326}:
464.

\bibitem{binder1981}K. Binder, Z. Phys. B, 1981,  {\bf 43}: 119.


\end{thebibliography}
\end{document}